\documentclass[lettersize,journal]{IEEEtran}
\IEEEoverridecommandlockouts
\usepackage{cite,comment}
\usepackage{algorithm}
\usepackage{algpseudocode}
\usepackage{amsmath,amssymb,amsfonts}
\usepackage{graphicx}
\usepackage{caption}
\usepackage{textcomp}
\usepackage{xcolor}
\usepackage{accents}
\def\BibTeX{{\rm B\kern-.05em{\sc i\kern-.025em b}\kern-.08em
    T\kern-.1667em\lower.7ex\hbox{E}\kern-.125emX}}
\begin{document}

\title{Prioritized Multi-Tenant Traffic Engineering for Dynamic QoS Provisioning in Autonomous SDN-OpenFlow Edge Networks}  

\author{\IEEEauthorblockN{Mohammad Sajid Shahriar, Faisal Ahmed, Genshe Chen, Khanh D. Pham, Suresh Subramaniam,\\ Motoharu Matsuura, Hiroshi Hasegawa, and Shih-Chun Lin}

\thanks{
Mohammad Sajid Shahriar, Faisal Ahmed, and Shih-Chun Lin are with North Carolina State University, NC, USA (e-mail: mshahri@ncsu.edu; fahmed5@ncsu.edu; slin23@ncsu.edu).

Genshe Chen is with Intelligent Fusion Technology, Inc., MD, USA (e-mail: gchen@intfusiontech.com).

Khanh D. Pham is with the Air Force Research Laboratory (AFRL), NM, USA (e-mail: khanh.pham.1@spaceforce.mil).

Suresh Subramaniam is with George Washington University, WA, USA (e-mail: suresh@gwu.edu).

Motoharu Matsuura is with the University of Electro-Communications, Tokyo, Japan (e-mail: m.matsuura@uec.ac.jp). 

Hiroshi Hasegawa is with Nagoya University, Nagoya, Japan (e-mail: hasegawa@nuee.nagoya-u.ac.jp).}
}

\maketitle

\begin{abstract}
This letter indicates the critical need for prioritized multi-tenant quality-of-service (QoS) management by emerging mobile edge systems, particularly for high-throughput beyond fifth-generation networks. Existing traffic engineering tools utilize complex functions baked into closed, proprietary infrastructures, largely limiting design flexibility, scalability, and adaptiveness. Hence, this study introduces a software-defined networking (SDN)-based dynamic QoS provisioning scheme that prioritizes multi-tenant network traffic while focusing on the base station-edge cloud scenario. The designed scheme first separates control and data planes and enables traffic management automation using SDN programmability. It then implements dynamic QoS management via the SDN-OpenFlow protocol, which ensures ample bandwidth for multiple priority flows and efficiently manages the remaining bandwidth for non-priority traffic. Empirical experiments are conducted with a Mininet network emulator and an OpenDayLight controller. Performance evaluation validates the proposed scheme's effectiveness in meeting multi-tenant QoS criteria, offering a robust solution for traffic prioritization in SDN-based edge networks. 
\end{abstract}

\begin{IEEEkeywords}
Multi-tenancy, prioritized traffic engineering, dynamic QoS provisioning, SDN-OpenFlow platforms, autonomous network management, and edge systems.
\end{IEEEkeywords}

\section{Introduction}

\IEEEPARstart{T}{he advent} of beyond-fifth-generation (B5G) mobile technology underscores the necessity for uniform control plane operations \cite{corici2023organic} and dynamic quality-of-service (QoS) management to meet its standards \cite{6gdt}, which aim for higher data rates and reduced latency. Traditional network infrastructures, however, face challenges in adhering to these standards due to their complexity and the limitations of existing QoS management protocols, such as DiffServ \cite{b1}, which are either too complex for practical use or lack the needed precision \cite{b2}. In response to these challenges, software-defined networking (SDN) has emerged as a significant innovation, providing a solution to the rigidity and complexity of conventional cellular networks by separating the control and data planes. This separation enhances network programmability and centralizes control, offering a holistic view of network resources, which facilitates improved traffic engineering and QoS management \cite{trafficeng}, \cite{b3}.

SDN's dynamic adaptability to changes in policy and configuration requirements marks a departure from the manual and static configurations of traditional networks. Its flexibility is particularly advantageous for traffic engineering and prioritization, offering a promising avenue for managing various applications, including unmanned aerial vehicles, autonomous vehicles, smart factories, and public safety operations \cite{b3}. A notable application of SDN is in prioritizing network traffic for emergency services, crucial for efficient relief and response during large public gatherings \cite{b5}. This requires adaptive network solutions capable of tailoring resource allocation to the specific needs of each emergency scenario, from high-bandwidth needs for quality video streaming to minimal bandwidth for basic communications. Guaranteeing bandwidth allocation for emergency or priority traffic while optimizing the use of remaining bandwidth for lower-priority traffic showcases SDN's potential to enhance network responsiveness and efficiency in critical situations.

In consideration of the foregoing observations, this letter introduces an SDN-based traffic engineering scheme that is designed to guarantee the allocation of ample bandwidth for emergency network traffic. Simultaneously, it optimizes the allocation of remaining network bandwidth to low-priority traffic flows through the utilization of the SDN-OpenFlow \cite{b6} protocol. The main contributions are summarized below: 

\begin{itemize}

\item To enhance low latency communication for multi-tenant environments, this study introduces an innovative SDN-based edge-cloud platform within the network architecture. This platform aims to minimize the distance between data sources and the network's base stations, facilitating reduced latency and intelligent networking solutions.

\item To prioritize the QoS provisioning in multi-tenant communication systems, this work utilizes a traffic classification method to identify priority traffic. Moreover, dynamic QoS provisioning is proposed to establish QoS goals and create rules to prevent any decline in priority traffic flow performance.

\item The effectiveness of the proposed dynamic QoS provisioning scheme is realized by utilizing an OpenDayLight (ODL)-based controller to collect traffic data and enforce QoS policies while simulating the data plane by using a Mininet network emulator to strategically control the forwarding and metering of flows.

\end{itemize}

\section{State-of-the-Arts}

Recent studies have emphasized integrating QoS within the SDN framework to enhance bandwidth guarantees and achieve general QoS objectives. While some research~\cite{b7,b8} have focused on QoS improvements, they have not specifically concentrated on bandwidth assurance. To bridge this gap, authors in \cite{b9} categorize network flows into QoS and best-effort flows, employing a path selection strategy that considers requested rates for improved bandwidth management. A significant development is considered in \cite{b10}, where OpenFlow meters and queues are used to differentiate flows, ensuring minimum bandwidth for QoS flows and integrating an admission control process to verify network path support for these flow rates. This bandwidth-guarantee design includes traffic flow monitoring at ingress switches and uses multi-queue configurations at egress ports for prioritization. Also, MPLS tunnels are implemented from bandwidth assurance~\cite{b2}; pre-configured network slices are provided for optimizing QoS and throughput~\cite{b11}. 

Despite these prior accomplishments, managing QoS in networks with multiple tenants' priority flows remains complex. Hence, this letter introduces a dynamic QoS provisioning scheme within an SDN-based traffic engineering platform aimed at prioritizing emergency traffic and enhancing resource allocation in multi-tenant scenarios, representing a significant advancement in network traffic management.

\setlength{\abovecaptionskip}{3px}

\setlength{\belowcaptionskip}{-2px}

\begin{figure}[t]
  \centering
    \includegraphics[width=3.4in]{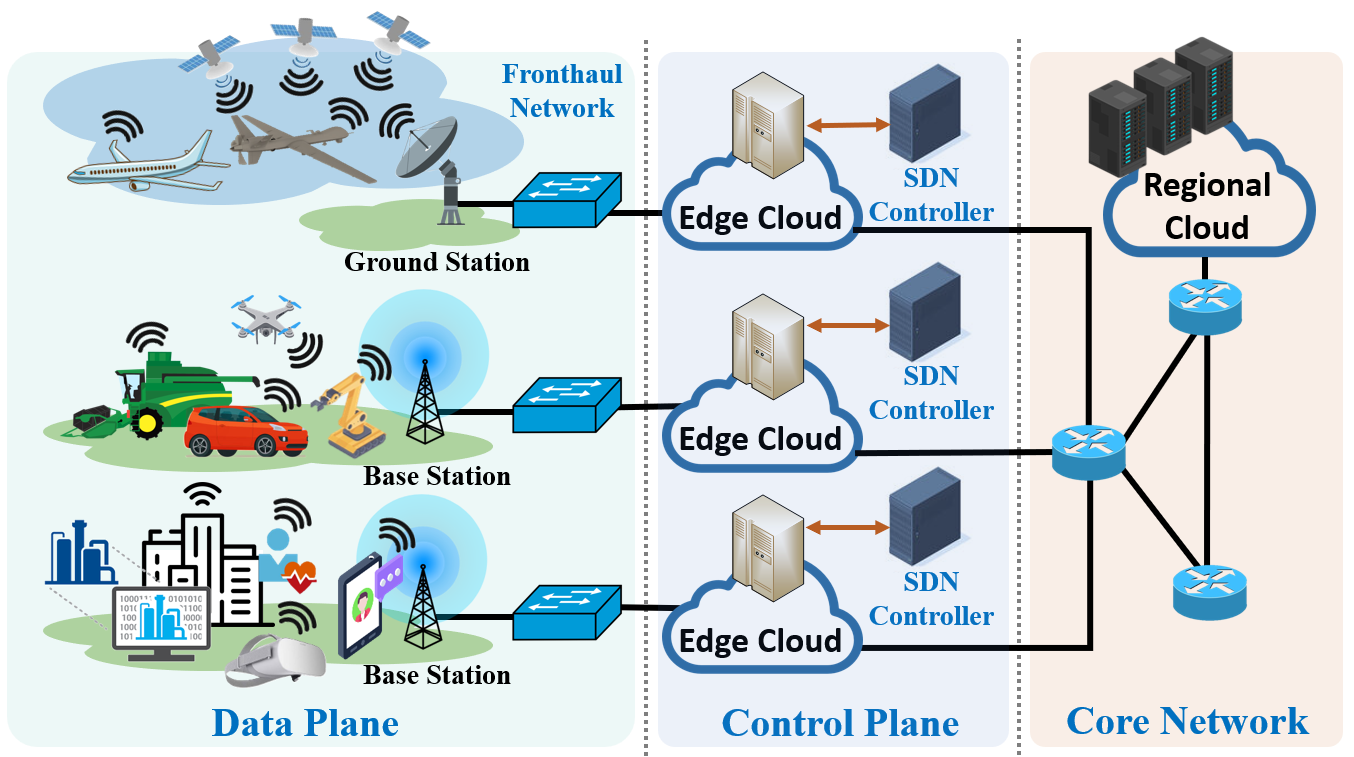} 
  \caption{An exemplary scenario of SDN-enabled multi-tenant edge networks.}
  \label{fig:scenario} 
\end{figure}

\section{Autonomous Software-Defined Edge Networks}
This paper introduces a novel approach to managing QoS effectively by prioritizing traffic, aiming to enhance open and smart networking in 6G communication systems. Fig.~\ref{fig:scenario} shows our considered SDN infrastructure in a BS-EC scenario (e.g., an SDN cellular system~\cite{softair}) that can accommodate multiple heterogeneous tenants. This system comprises a single region cloud (RC), $E$ number of ECs that represented by set $\mathsf{E} = \{1,2,3,\dots,e,\dots,E\}$, $N$ number of BSs, and $G$ number of ground stations (GSs) that indexed by set $\mathsf{N} = \{1,2,3,\dots,n,\dots,N\}$ and $\mathsf{G} = \{1,2,3,\dots,g,\dots,G\}$, respectively. Each EC is connected to an SDN controller, which orchestrates the data plane and the traffic flows between the EC and the tenants. For simplicity, we will focus on a single EC, denoted as $e$, and let $\mathsf{N}_e = \{1,2,3,\dots,n_{e},\dots,N_{e}\}$ and $\mathsf{G}_e = \{1,2,3,\dots,g_{e},\dots,G_{e}\}$ denote the set of BSs and GSs that are covered by that EC $e$, respectively.    

\begin{figure}[t]
  \centering
    \includegraphics[width=3.4in]{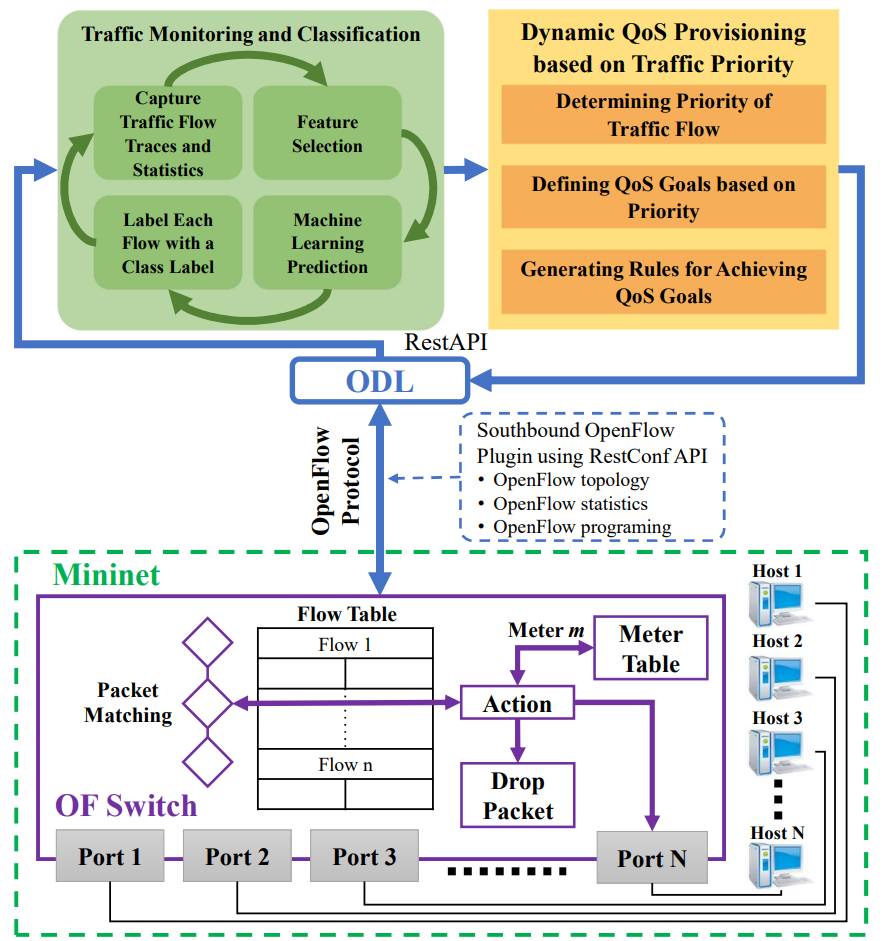} 
  \caption{The proposed prioritized multi-tenant traffic engineering platform.}
  \label{fig:systemmodel} 
\end{figure}

As illustrated in Fig. \ref{fig:systemmodel}, the operation of the proposed system commences with monitoring data traffic and their classification, subsequently translating these classifications into well-defined QoS objectives. After that, instructions for prioritizing bandwidth allocation are generated and employed by the data plane utilizing the ODL service functions.

The SDN framework presented encompasses a control plane responsible for managing flow tables and instructions and a data plane focused on enacting match-plus-action forwarding rules. It includes an ODL controller and a northbound application configuring QoS priorities via RestAPI, exploiting OpenFlow programming. Additionally, the setup integrates a simulated data plane using Mininet with OpenFlow switches for optimal flow forwarding and metering aimed at achieving the desired QoS. Mininet acts as a network emulator, facilitating the development and evaluation of complex network configurations on a single server.

\section{Multi-Tenant Quality-of-Service Provisioning with Traffic Prioritization}

To prioritize QoS provisioning for traffic originating from various tenants, our proposed system maintains a set of pre-established QoS goals. These goals outline the target bandwidth for each priority class, corresponding to the traffic generated by distinct tenants. Meter tables within the OpenFlow switches which work similar to leaky buckets, are used to put the QoS goals into effect. By configuring meter bands with the flow tables, we gain control over the bandwidth usage of priority and non-priority traffic flows. This requires distinguishing between priority and non-priority traffic, which is achieved by operating with a predefined set of network traffic classes \(C_p\). These are predefined as ``priorities," and each of them is associated with a specific QoS goal \(Q\). The system classifies a set of traffic classes \(C_t\) utilizing the traffic classification method (such as via a machine learning approach~\cite{b12}) by monitoring network traffic at a period of $T$. It then determines whether \(C_t\) is present in \(C_p\). Accordingly, our design tackles the prioritization problem in three scenarios:
\begin{itemize}
  \item A single match, (i.e., \(|C_p \cap C_t| = 1\));
  \item Multiple matches (i.e., \(|C_p \cap C_t| > 1\));
  \item No match of priority class (i.e., \(|C_p \cap C_t| = 0\)).
\end{itemize}
In the first and second cases, the goal is to provide better QoS to priority traffic. However, the second scenario poses the challenge of prioritizing bandwidth between multiple priority flows, assuming the maximum bandwidth of every link is $B_{\text{max}}$. However, where there is no match of a priority class, guaranteed QoS is not required. This scheme addresses the challenges of the scenarios by generating specific meter bands and QoS policies using \textbf{Algorithm 1}. It also considers the QoS for non-priority flows when the parameter $b_{\text{min}}$ is set to a value greater than zero. This parameter specifies the minimum bandwidth assigned to each non-priority flow when priority flows are present.

\begin{algorithm} [t] 
\caption{Generating QoS Rules Based on Traffic Priority.}
\begin{algorithmic}[1]
\If{\(|C_p \cap C_t| = 1\)  and  \(|C_t| > 1\)}
    \If{\(Q(C_p \cap C_t) > B_{\text{max}} - (|C_t| - 1)*b_{\text{min}}\)}
        \If{$b_{\text{min}} > 0$}
            \State Set meter of \( B_{\text{max}} - (|C_t| - 1)*b_{\text{min}} \) to flow \hspace*{1.5cm} \(C_p \cap C_t \)
            \State Set meter band of $b_{\text{min}}$ to the flows of \(C_t - C_p\)
        \Else
            \State Drop packet for the flows of \(C_t - C_p\)
        \EndIf
    \Else
        \If{$b_{\text{min}} > 0$}
            \State Set meter of \( Q(C_p \cap C_t) \) to the flows of $C_p \cap C_t$
        \EndIf
        \State Set meter of \(\frac{B_{\text{max}} - Q(C_p \cap C_t)}{|C_t| - 1}\) to flows of \(C_t - C_p\)
    \EndIf
\ElsIf{\(|C_p \cap C_t| > 1\)}
    \If{\(\sum_{d \in C_p \cap C_t} Q(d) > B_{\text{max}} - |C_t - C_p|*b_{\text{min}}\)}
        \State \( r = \frac{B_{\text{max}} - |C_t - C_p|*b_{\text{min}} }{\sum_{d \in C_p \cap C_t} Q(d)}\) 
        \State Set meter band of $r * Q(d)$ to flows of \(C_p \cap C_t\)
        \If{$b_{\text{min}} > 0$}
            \State Set meter band of $b_{\text{min}}$ to flows \(C_t - C_p\)
        \Else
            \State Drop packet for the flows of \(C_t - C_p\)
        \EndIf
    \Else
        \State Set meter band of $Q(c)$ to flows of \(C_p \cap C_t\)
        \State Set meter of \(\frac{B_{\text{max}} - \sum_{d \in C_p \cap C_t} Q(d)}{|C_t - C_p|}\) to flows \(C_t - C_p\)
    \EndIf
    
\Else
    \State Reset the flow table to default 
\EndIf
\end{algorithmic}
\end{algorithm}

In \textbf{Algorithm 1}, our proposed traffic engineering scheme adjusts meter bands and updates flow entries when dealing with one or multiple priority classes. The use cases are outlined below:
\begin{itemize}
 
  \item {\textbf{Case 1 ({\textit{Single-Priority Flow})}}}: In cases where the network traffic includes a single priority flow and at least one non-priority flow, \textbf{Algorithm 1} makes strategic decisions on packet dropping using designated or calculated meter bands. The choice of imposing meter on either both priority and non-priority flows or solely on non-priority flows depends on whether the combined bandwidth goals for priority and non-priority flows exceed $B_{\text{max}}$ and if the value of $b_{\text{min}}$ is greater than zero.

  \item {\textbf{Case 2 (\textit{Multi-Priority Flow})}}: In more complex scenarios featuring multiple-priority classes within the network traffic, if the total target bandwidth for all priority flows (i.e., represented by $\sum_{d \in C_p \cap C_t} Q(d)$) exceeds $B_{\text{max}}$, the dynamic QoS provisioning of the scheme carefully adjusts the meter rates for each priority flow. This adjustment is made using a factor $r$ that considers the $Q$ value of each priority flow. Conversely, when the total target bandwidth is within $B_{\text{max}}$, the scheme utilizes the $Q$ values to set meter rates for each priority flow, ensuring they do not compete for bandwidth among themselves. Simultaneously, the scheme considers non-priority flows based on $b_{\text{min}}$, similar to the previous use case.
  
  \item {\textbf{Case 3 (\textit{No-Priority Flow})}}: In instances where the network traffic lacks any priority flow, the scheme reverts the flow entries to their initial configuration. This reset signifies a return to a baseline state, where no specific traffic class is given preferential treatment in terms of bandwidth allocation.
  
\end{itemize}

\section{Performance Evaluation}
\begin{figure}[t]
  \centering
    \includegraphics[width=3.4in]{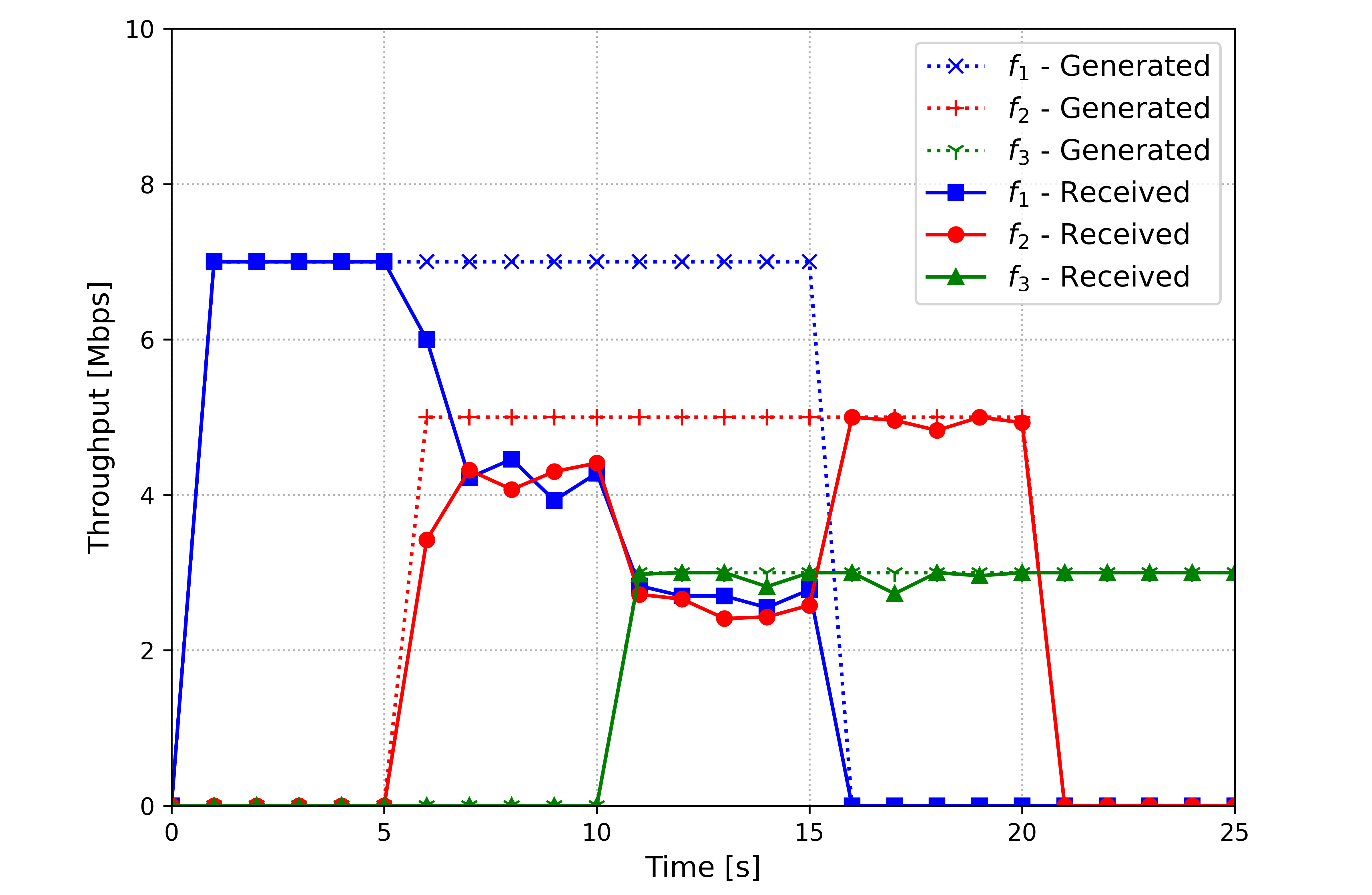} 
  \caption{Throughput of individual flows without employing the proposed traffic engineering platform.}
  \label{fig:result1} 
\end{figure}

\begin{figure}[t]
  \centering
    \includegraphics[width=3.4in]{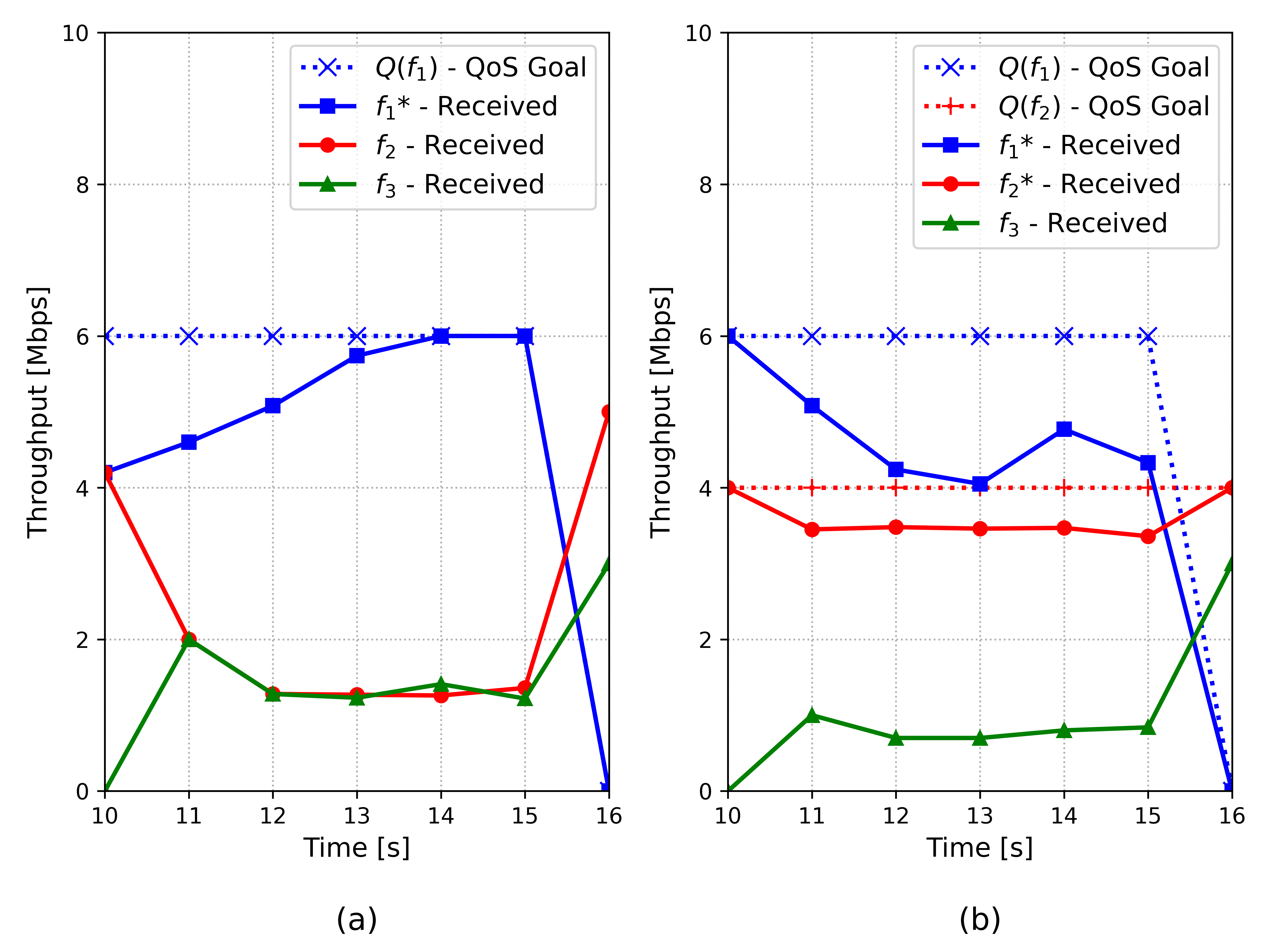} 
  \caption{Throughput of each flow in (a) single-priority and (b) multi-priority cases (priority flows are marked with `*') under the provisioning of proposed traffic engineering platform.}
  \label{fig:result2} 
\end{figure}

To assess the effectiveness of the proposed scheme, we analyze three flows ($f_1$, $f_2$, and $f_3$) using Mininet simulation, as illustrated in Fig. \ref{fig:systemmodel}. It is assumed that $f_1$, $f_2$, and $f_3$ correspond to base stations $BS_1$, $BS_2$, and $BS_3$, respectively. 
Figs.~\ref{fig:result1} and \ref{fig:result2} show the generated and received traffic rates over a 25-second period, with every $T=5$ second for flow priority identification by the SDN controller. However, fig. \ref{fig:result2} emphasizes on specific time period to demonstrate the advantage of the proposed scheme. The traffic patterns for $f_1$, $f_2$, and $f_3$ are configured at $7 Mbps$, $5 Mbps$, and $3 Mbps$, respectively, to demonstrate various scenarios—such as with one, multiple, or no priority flows—where the overall traffic might exceed the total link capacity $B_{\text{max}}$ set at $10 Mbps$.

In Fig. \ref{fig:result1}, the proposed autonomous edge network and QoS provisioning scheme are not utilized. It reveals that flows from different tenants compete for bandwidth, resulting in throughput degradation when the overall generated traffic surpasses the link bandwidth. 

Fig. \ref{fig:result2} (a) demonstrates that for case 1, among the three flows, only $f_1$ (marked by *) is identified as the priority flow with a $Q$ value of $6 Mbps$ (i.e., the suggested throughput). The drop meters set by the proposed scheme effectively limit the traffic rate of non-priority flows, preventing interference in achieving the target throughput of the priority flow. Fig. \ref{fig:result2} (b) shows the outcomes when the traffic engineering platform recognizes two priority flows, $f_1$ and $f_2$ (marked by *), each having $Q$ values of $6 Mbps$ and $4 Mbps$, respectively. This highlights the influence of drop meters dynamically configured by the scheme at regular intervals to handle the $b_{\text{min}}$ of $1 Mbps$ for the non-priority flow. The result demonstrates that the proposed scheme not only prevents major degradation in performance for priority flows but also accommodates the non-priority flow, ensuring that the minimum QoS is maintained. However, the data transmissions in Mininet do not always reach the expected bandwidth because of overheads and shared resources among hosts.

\vspace{-10px}

\section{Conclusions}

We introduced novel dynamic QoS provisioning within an SDN edge framework tailored for prioritizing traffic in multi-tenant networks. The proposed scheme revamped the SDN-OpenFlow protocol and ensured QoS adherence by allocating specific bandwidth employing meters bands in the data plane for precise data flow control, effectively managing bandwidth to enable target QoS achievement. Empirical experiments validated our design's efficacy in bandwidth allocation, multi-tenancy sharing, and QoS policy management, thus setting this work as a reference for future traffic engineering.

\section*{Acknowledgment}
This material is partially based upon work supported by the United States Air Force under contract No. FA9453-23-P-A044. Any opinions, findings and conclusions or recommendations expressed in this material are those of the authors and don’t necessarily reflect the views of the AFRL, United States Air Force, or the U.S. Government. 
We gratefully acknowledge support from the AFRL, the NC Space Grant, the National Science Foundation (NSF) under Grant CNS-2210344, the North Carolina Department of Transportation (NCDOT) under Award TCE2020-03, and Meta 2022 AI4AI Research.


\end{document}